\documentclass[floatfix,%
 reprint,
superscriptaddress,
nofootinbib,
 amsmath,amssymb,
 aps,
prd,
]{revtex4-2}

\usepackage{ragged2e}
\usepackage{graphicx}
\usepackage{caption}
\usepackage{subcaption}
\usepackage{dcolumn}
\usepackage{bm}
\usepackage{hyperref}
\usepackage{multirow}

\newcommand{\be}{\begin{equation}}
\newcommand{\ee}{\end{equation}}
\newcommand{\app}{\texttt{NRHybSur3dq8} }
\newcommand{\dan}{\texttt{DANSur3dq8} }
\newcommand{\aes}{\texttt{AESur3dq8} }
\newcommand{\qcbbh}{quasi-circular binary-black-hole mergers }
\newcommand{\gwsur}{\textsc{gwsurrogate} }

\usepackage{multibib}

\usepackage[dvipsnames]{xcolor}

\DeclareCaptionJustification{justified}{\justifying}
\usepackage[normalem]{ulem} 

\begin{document}


\title{An autoencoder-based surrogate waveform model for quasi-circular binary-black-hole mergers}

\author{Anastasios Theodoropoulos}
\affiliation{Departamento de
  Astronom\'{\i}a y Astrof\'{\i}sica, Universitat de Val\`encia,
  Av.~Vicent Andrés Estellés 19, 46100, Burjassot (Val\`encia), Spain}
\email{anastasios.theodoropoulos@uv.es}

\author{Nino Villanueva}
\affiliation{IDAL, Electronic Engineering Department, ETSE-UV, University of Valencia, Avgda. Universitat s/n, 46100 Burjassot, Valencia, Spain}
\affiliation{Departamento de
  Astronom\'{\i}a y Astrof\'{\i}sica, Universitat de Val\`encia,
  Av.~Vicent Andrés Estellés 19, 46100, Burjassot (Val\`encia), Spain}
\author{Osvaldo~Gramaxo Freitas}
\affiliation{Centro de F\'{\i}sica das Universidades do Minho e do Porto (CF-UM-UP), Universidade do Minho, 4710--057 Braga, Portugal}
\affiliation{Departamento de
  Astronom\'{\i}a y Astrof\'{\i}sica, Universitat de Val\`encia,
  Av.~Vicent Andrés Estellés 19, 46100, Burjassot (Val\`encia), Spain}


\author{Tiago Fernandes}
\affiliation{Centro de F\'{\i}sica das Universidades do Minho e do Porto (CF-UM-UP), Universidade do Minho, 4710--057 Braga, Portugal}
\affiliation{Departamento de
  Astronom\'{\i}a y Astrof\'{\i}sica, Universitat de Val\`encia,
  Av.~Vicent Andrés Estellés 19, 46100, Burjassot (Val\`encia), Spain}

\author{Solange Nunes}
\affiliation{Centro de F\'{\i}sica das Universidades do Minho e do Porto (CF-UM-UP), Universidade do Minho, 4710--057 Braga, Portugal}
\author{Alejandro~\surname{Torres-Forn\'e}}
\affiliation{Departamento de
  Astronom\'{\i}a y Astrof\'{\i}sica, Universitat de Val\`encia,
  Av.~Vicent Andrés Estellés 19, 46100, Burjassot (Val\`encia), Spain}
\affiliation{Observatori Astron\`omic, Universitat de Val\`encia,  Catedr\'atico 
  Jos\'e Beltr\'an 2, 46980, Paterna (Val\`encia), Spain}

\author{Jos\'e~A.~Font}
\affiliation{Departamento de
  Astronom\'{\i}a y Astrof\'{\i}sica, Universitat de Val\`encia,
  Av.~Vicent Andrés Estellés 19, 46100, Burjassot (Val\`encia), Spain}
\affiliation{Observatori Astron\`omic, Universitat de Val\`encia,  Catedr\'atico 
  Jos\'e Beltr\'an 2, 46980, Paterna (Val\`encia), Spain}

\author{Antonio~Onofre}
\affiliation{Centro de F\'{\i}sica das Universidades do Minho e do Porto (CF-UM-UP), Universidade do Minho, 4710--057 Braga, Portugal}

\author{Jos\'e D. Mart\'in-Guerrero}
\affiliation{IDAL, Electronic Engineering Department, ETSE-UV, University of Valencia, Avgda. Universitat s/n, 46100 Burjassot, Valencia, Spain}
\affiliation{Valencian Graduate School and Research Network of Artificial Intelligence (ValgrAI), Spain}

\date{\today}

\begin{abstract}
The generation of accurate waveforms from binary black hole (BBH) mergers is a major effort in Gravitational-Wave Astronomy. In recent years, machine-learning–based surrogate models for BBH waveforms have been proposed. Those offer the potential to dramatically accelerate waveform generation while maintaining accuracy competitive with that of traditional waveform approximants. In this work, we investigate the viability of autoencoders as generative models for gravitational-wave signals from quasi-circular BBH mergers. We introduce \texttt{AESur3dq8}, a novel surrogate waveform model based on autoencoders that enables the rapid and accurate construction of large template banks, producing millions of waveforms in under a second using modest computational resources. The model is trained on the numerical-relativity–informed surrogate \app and subsequently fine-tuned using the SXS catalog of BBH simulations. We demonstrate that waveforms generated by \texttt{AESur3dq8} achieve mismatches of order $10^{-4}$ with respect to Numerical Relativity waveforms, and that parameter estimation performed with these templates yields results fully consistent with those reported by the LIGO–Virgo–KAGRA Collaboration for observed gravitational-wave events.
\end{abstract}

\maketitle


\section{\label{sec:intro} Introduction}

Gravitational waves offer a brand-new window to the physical world, complementing the wealth of information collected through the electromagnetic channel. In particular, they provide a unique opportunity to probe gravity and matter in some of the most extreme conditions found in Nature. Their long-awaited detection was made possible by the LIGO \cite{LIGOScientific:2016aoc} and Virgo \cite{VIRGO:2014yos} detectors. The latest release of the Gravitational-Wave Transient Catalog from the LIGO-Virgo-KAGRA (LVK) collaboration has recently updated the number of confident detections to 218 observations, all of them associated with compact binary coalescences, predominantly binary black hole (BBH) mergers~\cite{abbott_gwtc-1_2019,abbott_gwtc-2_2021,abbott_gwtc-3_2021,abbott_gwtc-21_2024,LIGOScientific:2025jau}. As the sensitivity of current detectors is improved \cite{collaboration_advanced_2015,VIRGO:2014yos} and new detectors join the observations~\cite{hall_cosmic_2022,ET:2019dnz,ET:2025xjr}, the amount of gravitational-wave observations is expected to significantly grow. New facilities as the Einstein Telescope or the Cosmic Explorer are expected to achieve detection rates of order 100,000 BBH mergers per year~\cite{ET_rates,hall_cosmic_2022}. As a result, accelerating both the detection and parameter estimation capabilities of current pipelines will soon become critical in order to extract accurate results for as many gravitational-wave signals as possible.  

Advances will undoubtedly hinge on the efficient and rapid processing of those data, turning Artificial Intelligence and Machine Learning (ML) into key methodological approaches to boost results. Their use, which is already substantial and keeps steadily growing (see~\cite{Cuoco:2020,Benedetto:2023,Stergioulas:2024,Cuoco:2025} for recent reviews), is likely going to widespread across all subareas of Gravitational-Wave Astronomy, from the mitigation of the noise of the detectors to the modelling and generation of waveforms of compact binary coalescences, from strategies for detection searches to source classification and astrophysical inferences. In this work, we focus on the problem of gravitational-waveform generation from compact binary coalescences, specifically BBH mergers. The accurate computation of such waveforms is computationally demanding, primarily due to the intrinsic difficulty of solving the equations of motion governing binary systems in General Relativity, whether analytically or numerically. Analytic approaches lose accuracy in the late inspiral and merger phases of compact binary coalescences, where strong-field and tidal effects become dominant, leaving Numerical Relativity as the only viable method for first-principles waveform computation. Although Numerical Relativity has achieved remarkable success, it remains severely limited in its ability to model generic binary configurations, particularly those involving large mass ratios, misaligned spins, or non–quasi-circular (eccentric) orbits. Even for relatively simple systems, Numerical Relativity simulations require substantial computational resources, restricting current catalogs to approximately ${\cal O}(10^3)$ numerically generated waveforms for BBH systems, and roughly an order of magnitude fewer for binary neutron star systems. In contrast, gravitational-wave searches conducted by the LVK Collaboration typically rely on template banks containing ${\cal O}(10^6-10^7)$ waveforms. The vast majority of these are generated using so-called approximants, i.e.~fast, semi-analytic waveform models that produce accurate predictions as functions of the source’s physical parameters (see~\cite{LISAConsortiumWaveformWorkingGroup:2023arg} for a recent overview). These approximants are commonly calibrated against Numerical Relativity waveform catalogues and have enabled the successful analysis of all gravitational-wave signals detected to date. 

More specifically, this investigation focuses on the use of ML models to accelerate the gravitational-waveform generation process. Many of the surrogate models used for waveform generation rely on a reduced-basis representation of the waveform space, typically obtained through Principal Component Analysis (PCA) or Singular Value Decomposition (SVD), combined with an interpolation scheme over the physical parameters such as Gaussian Process Regression (GPR). In this context, GPR has been  employed to construct surrogate waveform models for both non-precessing and precessing BBH systems, enabling accurate predictions at regions of parameter space not directly covered by Numerical Relativity~\cite{Williams:2020}. A further example is provided by our previous work~\cite{GramaxoFreitas:2024bpk}, in which we introduced a neural-network–based surrogate model, heavily reliant on Principal Component Analysis (PCA), designed to rapidly and accurately generate non-precessing BBH waveforms. When applied to waveform approximant data, this model achieves median reconstruction mismatches of order $10^{-6}$ to $10^{-4}$, depending on the approximant. Correspondingly, median mismatches of order $10^{-5}$ are obtained when trained on Numerical Relativity datasets. The slight degradation in the latter case is primarily attributable to the limited availability of numerical waveforms for training. Performance benchmarks demonstrate that the PCA-based model of~\cite{GramaxoFreitas:2024bpk} can generate more than one million waveforms in less than 3 milliseconds when deployed on a GPU. When we re-tested on our RTX 4090 using sequential batching in order to accommodate VRAM constraints, it takes more time ($ 172$ ms). It is still faster than our autoencoder-based model because of its simplicity and lower size. Recent studies have further explored the potential of neural networks to enhance the efficiency of surrogate waveform models. For instance, \cite{Fragkouli:2022} implemented a four-layer neural network to predict the interpolation coefficients in the parameterised surrogate framework originally introduced in~\cite{Field:2014}. Similarly, \cite{nousi_autoencoder-driven_2022} adopted the same surrogate methodology but replaced traditional regression techniques with autoencoders. Their analysis uncovered an intriguing ``spiral" structure in the latent space, reflecting the dependence of the fitting coefficients on the mass ratio. By exploiting the autoencoder's ability to learn smooth, differentiable mappings between input parameters and surrogate coefficients, the authors achieved faster and more accurate regression. A comprehensive overview of these developments can be found in~\cite{Stergioulas:2024}.

Here, we build on our previous investigation \cite{GramaxoFreitas:2024bpk}, developing a new surrogate waveform model dubbed \texttt{AESur3dq8} based on autoencoders, which are a generalisation of PCA. A previous work employing autoencoders~\cite{nousi_autoencoder-driven_2022} focused on non-spinning BBH mergers, for which the mass ratio is the sole input parameter. In this framework, the autoencoders are used primarily as a representation-learning tool. They learn a low-dimensional latent encoding of the waveform data, which subsequently informs a fully connected neural network responsible for generating the final waveform. Both PCA and autoencoders capture the spiral structure found in the coefficient space, with autoencoders providing a slightly better representation.  We extend the approach of~\cite{nousi_autoencoder-driven_2022} by using the autoencoder not merely as a representation-learning tool but, through the introduction of an additional fully connected network upstream of the decoder, as a hybrid artificial neural network capable of producing gravitational-wave signals from quasi-circular BBH systems. Moreover, we further extend the model of~\cite{nousi_autoencoder-driven_2022} by incorporating the individual black hole spins as explicit input parameters of the network, in addition to the mass ratio.

The structure of this paper is the following: In Section~\ref{sec:dataset} we present the methodology to build our datasets. Section~\ref{sec:methods} describes our ML model, its specific architectures and the training procedure. In Section \ref{sec:res} we discuss the main results of our analysis, namely the accuracy of the generated waveforms and the speed of the generation process as a function of the samples used. Moreover, this section also reports on the parameter estimation of an actual gravitational-wave signal using the \textsc{bilby} pipeline~\cite{ashton_bilby_2019}, comparing the posterior distributions of the BBH parameters obtained using our surrogate waveform and a baseline approximant. Finally, Section \ref{sec:con} presents our conclusions. The surrogate waveform models produced in this work are publicly available\footnote{https://github.com/TasosTheodoropoulos/AESur3dq8.git}.

\section{\label{sec:dataset} Datasets}

Deep Learning frameworks usually require large datasets in order to properly interpolate between the input and output parameter spaces. This becomes a major problem when applying Deep Learning methods in the context of Numerical Relativity, as current catalogs contain at most only a few thousand waveforms~\cite{Mroue:2013xna,Boyle:2019kee,Scheel:2025jct,Healy:2020vre,Healy:2022wdn,Kiuchi:2017pte,Kuan:2025bzu}. Our approach to this problem is to generate a substantial dataset using an existing surrogate model that contains most of the information about what a gravitational-wave signal should look like and then fine-tune the model on our smaller dataset consisting of the Numerical Relativity produced waveforms. This is a widely used technique in the context of object recognition~\cite{li2023aligndet}, classification, as well as Large Language Models~\cite{parthasarathy2024ultimate}. In this work we focus on BBH coalescence signals, generating the approximant waveforms using the \app \cite{Varma:2018mmi,varma_surrogate_2019} surrogate model. The catalog we use is the SXS catalog\footnote{\href{https://data.black-holes.org}{data.black-holes.org}}, which is publicly available and features a relatively large number of waveforms in the parameter space we want to explore. 

\begin{figure}
    \centering
    \includegraphics[width=1\linewidth]{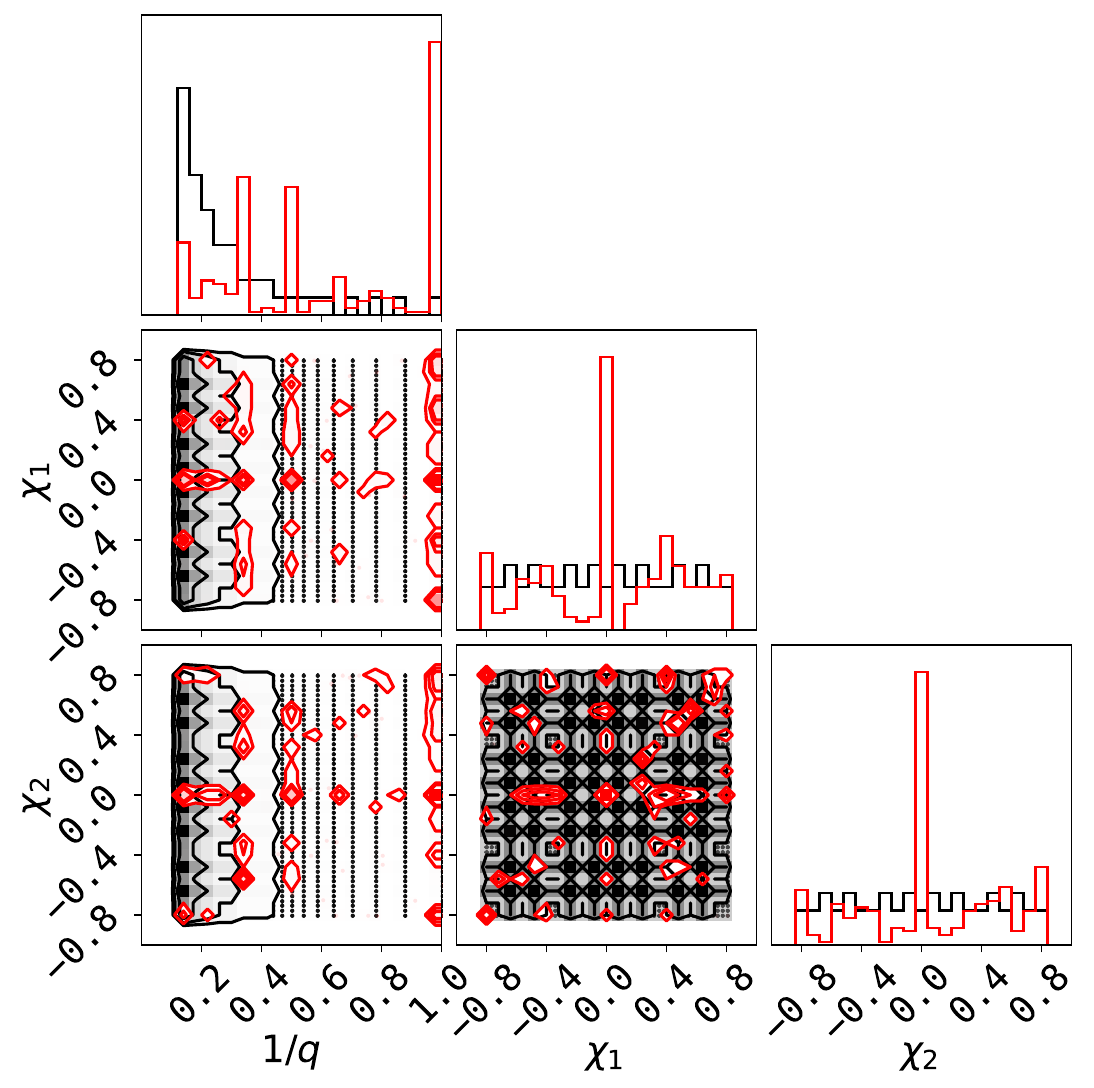}
    \caption{Corner plot showing the distribution of the parameters of the \app dataset, in gray and overlayed on top of it, in red, the respective distribution for the SXS dataset. The parameters, from top to bottom and left to right are, the inverse of the mass ratio $1/q$, the spin of the first black hole $\chi_1$ and of the second one $\chi_2$. The contour plots display the 0.5, 1, 1.5 and 2$\sigma$ isocontours, overlayed on top of a scatter plot of the parameters to show the density of points. Note that the points in the regions in-between isocontour levels are not displayed for clarity.}
    \label{fig:corner_plot_params}
\end{figure}

\subsection{Approximant dataset}

Using the \textsc{gwsurrogate}~\cite{Field:2025isp} python package, we generate 125,000 BBH waveforms for the \app model. Since we do not want to create an overly complex dataset that will make it difficult for the model to converge, we  uniformly sample both the BBH mass ratio $q=M_2/M_1$ in the interval [1,8] and the black hole individual spins $\chi_1, \chi_2$, in the interval [-0.8, 0.8]. Those are assumed to be aligned with the orbital angular momentum, since we only consider non-precessing, quasi-circular BBH mergers in this work. We note that we uniformly sample in mass ratios and spins in order to avoid biases by oversampling specific regions of the parameter space. 
The distributions of the parameters are displayed in Fig.~\ref{fig:corner_plot_params}. 

In a more technical note, the waveforms are generated in simulation units; the amplitudes/phases are scaled by the sum of the Christodoulou masses of both black holes and the sampling rate used is $2M_\odot^{-1}$. At the time of BBH merger we use the time corresponding to the maximum amplitude of the quadrupolar gravitational-wave mode $(l,|m|) = (2,2)$ and we line up the waveforms such that the merger is located in time $100M_\odot$ before the end of the window, which we choose to have a duration of $4096M_\odot$. Lastly, we further simplify the problem by setting the initial phase of all waveforms to zero. Using the \texttt{train\_test\_split} function~\cite{scikit-learn} and a 81/10/9  train/test/validation split on our dataset, we obtain the sets used for training, testing and validation of our network. This results in sets with 101250, 12500 and 11250
waveforms, respectively.

\begin{table}[htbp]
\centering
\caption{Detailed neural network architecture for the encoder, decoder and mapping network, containing the type, number of neurons and activation function for each layer. 
}
\label{tab:architecture_details}
\begin{tabular}{llll}
\toprule
\textbf{Component} & \textbf{Layer} & \textbf{Units} & \textbf{Activation} \\
\hline

\multirow{6}{*}{Encoder} 
& Input & 8192 & --- \\
& Dense 1 & 1024 & Leaky ReLU \\
& Dense 2 & 512 & Leaky ReLU \\
& Dense 3 & 256 & Leaky ReLU \\
& Dense 4 & 128 & Leaky ReLU \\
& Output (Latent) & 64 & Linear \\
\hline
\multirow{6}{*}{Decoder}
& Input (Latent) & 64 & --- \\
& Dense 1 & 128 & Leaky ReLU \\
& Dense 2 & 256 & Leaky ReLU \\
& Dense 3 & 512 & Leaky ReLU \\
& Dense 4 & 1024 & Leaky ReLU \\
& Output & 8192 & Linear \\
\hline
\multirow{6}{*}{Mapping Network}
& Input (Parameters) & 3 & --- \\
& Dense 1 & 128 & Leaky ReLU \\
& Dense 2 & 512 & Leaky ReLU \\
& Dense 3 & 512 & Leaky ReLU \\
& Dense 4 & 128 & Leaky ReLU \\
& Output (Latent) & 64 & Linear \\
\hline
\hline
\end{tabular}
\end{table}

\subsection{Numerical Relativity waveforms}

Using the publicly available BBH simulation dataset from the SXS collaboration~\cite{Scheel:2025}, which contains 2018 simulations, we extract only the $l=m=2$ modes with the \textsc{sxs} python package~\cite{sxs_package_2024}. Then, we filter these waveforms to make sure they match the parameters of our approximant dataset. We first make sure that we use waveforms that are longer than $4096M_\odot$. Then, we filter out all the waveforms with an effective precessing spin $\chi_p>0.001$, and mass ratios $q>8$. Lastly, we filter waveforms that contain large numerical errors, by filtering the total sum, as well as making sure the final amplitude $(A)$, after merger is adequately small. This is done by discarding waveforms with: $\sum_k A_k > 2000$ or $A_{\rm final} > 0.5$. This filter takes out only one problematic signal and we end up with 253 waveforms which we interpolate using cubic splines and trim  to the desired length ($4096M_\odot$). The parameter distributions for both datasets are shown in Fig.~\ref{fig:corner_plot_params}. We split our numerical relativity dataset in the same way we did for the \app dataset, resulting in train/test/validation sets with 204, 26 and 23 numerical waveforms, respectively.

\section{\label{sec:methods} Methodology}

The methodology we employ to generate new waveforms from the initial BBH parameters includes a mix of neural network architectures and training processes which we present in the following subsections.

\begin{figure}
     \centering
     \begin{subfigure}[b]{0.45\textwidth}
         \raggedleft    
         \includegraphics[width=0.9\textwidth]{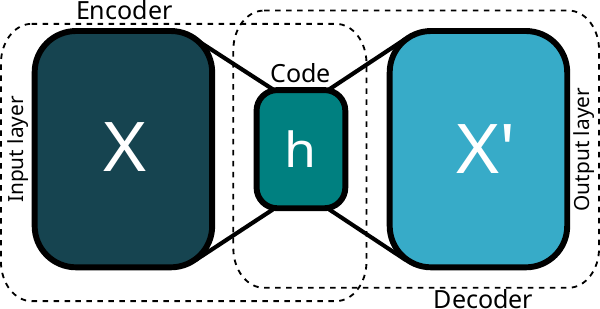}
         \caption{}
         \label{fig:autoencoder}
     \end{subfigure}
     \hfill
     \begin{subfigure}[b]{0.45\textwidth}
         \raggedleft
         \includegraphics[width=0.9\textwidth]{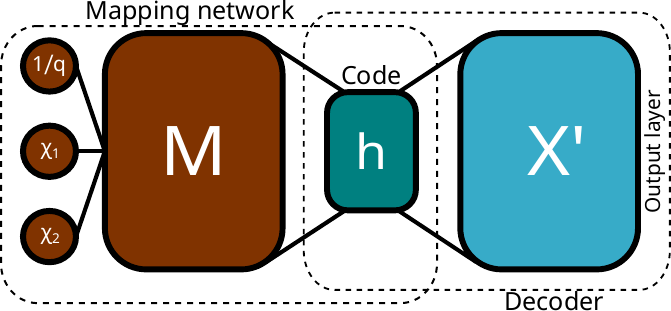}
         \caption{}
         \label{fig:full_model}
     \end{subfigure}
        \caption{Schematic representations of the two model configurations used in this work. (a) The autoencoder architecture, consisting of the encoder, decoder, and the latent (code) space. (b) The generative model, obtained by removing the encoder and adding a fully connected network that maps the physical parameters to the latent space.}
        \label{fig: Schematics}
\end{figure}

\begin{figure*}
    \includegraphics[width=1\linewidth]{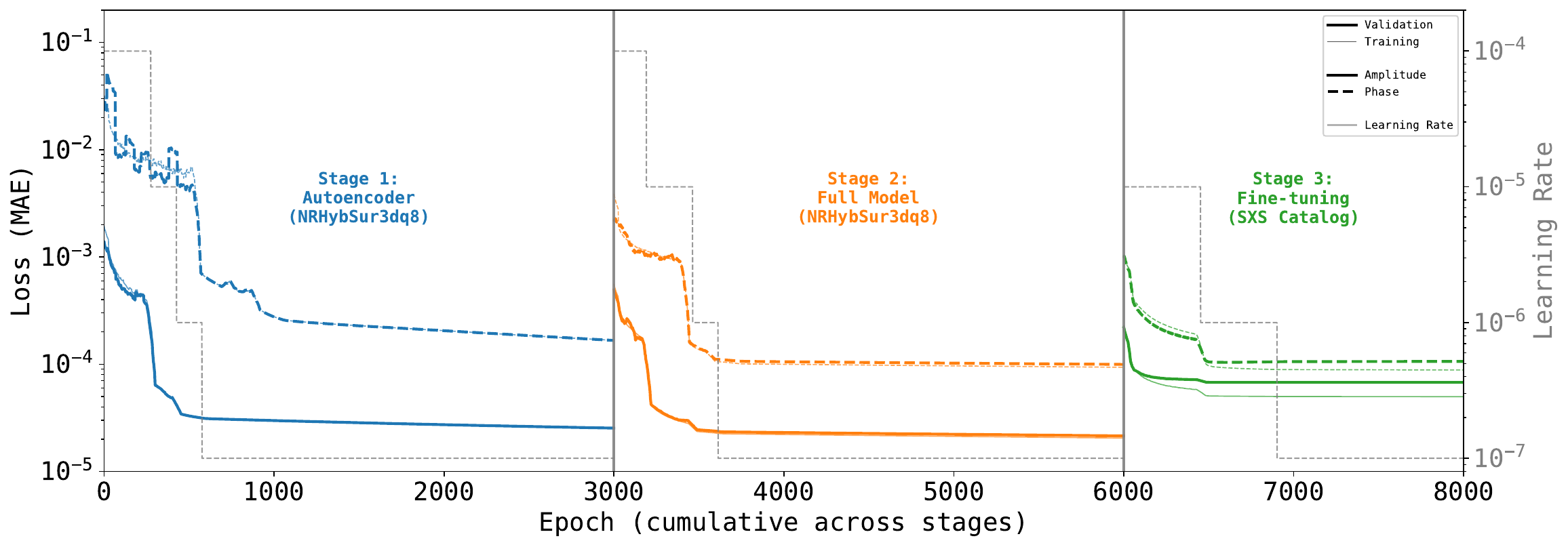}
    \caption{Loss functions of our models in all three stages of training. From left to right we show the 50-epoch average of the value of the loss and validation loss for each model (amplitude and phase, depicted in solid and dashed lines, respectively) during the autoencoder training with the \app dataset, the full model training with the same dataset, and the full model training with the SXS dataset (fine-tuning stage). We also overlay the LR for each epoch, showing more clearly that the model benefits from the LR scheduler, since the loss function improves at each step.}
    \label{fig:loss_hist}
\end{figure*}

\subsection{Dimensionality reduction}

The need for dimensionality reduction techniques comes from what is commonly referred to as the `curse of dimensionality', which refers to the observation that data get sparser in higher dimensions~\cite{altman_curses_2018}. As a result, models tend to overfit the data in order to achieve better accuracy or they tend to have redundant and/or correlated variables~\cite{noauthor_what_2024,noauthor_introduction_nodate}. To circumvent this issue one can either find more data so that a bigger part of the sparse regions can be covered, select some of the features of the original data, or even extract the most relevant features. In our case, since Numerical Relativity simulations of BBH mergers are computationally intensive, we cannot follow the first option. Moreover, as we want our model to be as accurate as possible we cannot use the second option either. Thus, we employ feature extraction techniques. Some commonly used such techniques are PCA and autoencoders. In~\cite{GramaxoFreitas:2024bpk} we focused on the generation of surrogate BBH waveforms using PCA. Here, we explore the use of autoencoders as the natural nonlinear generalisation of our previous work.

\subsection{Architectures}

An autoencoder is an unsupervised learning method, which uses a ``bottleneck'' architecture, as seen schematically in Fig.~\ref{fig:autoencoder}, where the input parameters get gradually encoded to less and less parameters until a preferable ``compression'' or encoding of the input data is reached. Then, a symmetric architecture is employed as the decoder network to ``decompress'' the encoded version of our data back to their original form. In this work, we adopted a 64-dimensional latent space, corresponding to a compression factor of 128. Our tests indicated that this choice provides the best trade-off between high reconstruction accuracy while avoiding overfitting to the waveform approximant dataset. We also employed an $\mathcal{L}_2$-norm \texttt{ activity\_regulariser} in the latent space layer which improved our overall results and helped the mapping network converge to more accurate results.

The detailed architectures of the encoder, decoder and mapping network are reported on Table~\ref{tab:architecture_details}. All the hidden layers use the \texttt{LeakyReLU} activation function, defined as
\begin{equation}
f(x) = \Biggl\{
    \begin{array}{lr}
        x, & \text{if } x \geq 0\\
        \\
        \alpha x, & \text{otherwise}
    \end{array}
\end{equation}
where $\alpha \in (0,1).$ For $\alpha =0$ the \texttt{LeakyReLU} reduces to the \texttt{ReLU} activation function, while in our case we use the default value $\alpha = 0.2$. On the other hand, for the input, output and latent layers we employ a linear activation function, $f(x) = x$. To convert the autoencoder into a generative model, we replace the encoder with a fully connected mapping network that projects the initial parameters into the latent space. These latent representations are then used as inputs to the decoder to generate the corresponding gravitational waveforms. The final architecture is shown in Fig.~\ref{fig:full_model}.

The loss function we use is the Mean Absolute Error (MAE),
\be
\mathcal{L} = \frac{1}{NL}\sum_{i}^N\sum_{j}^L \left( \left|X^i_{\rm true, j} - X^i_{\rm predicted, j}\right|\right)\,,
\ee
where $X_{\rm predicted}^i$ is the  waveform output of the model, $X_{\rm true}^i$ is the true value from the training dataset, and the difference of the two is averaged over the batch size $N$ and along their corresponding length $L$.

\begin{figure*}
    \includegraphics[width=1\linewidth]{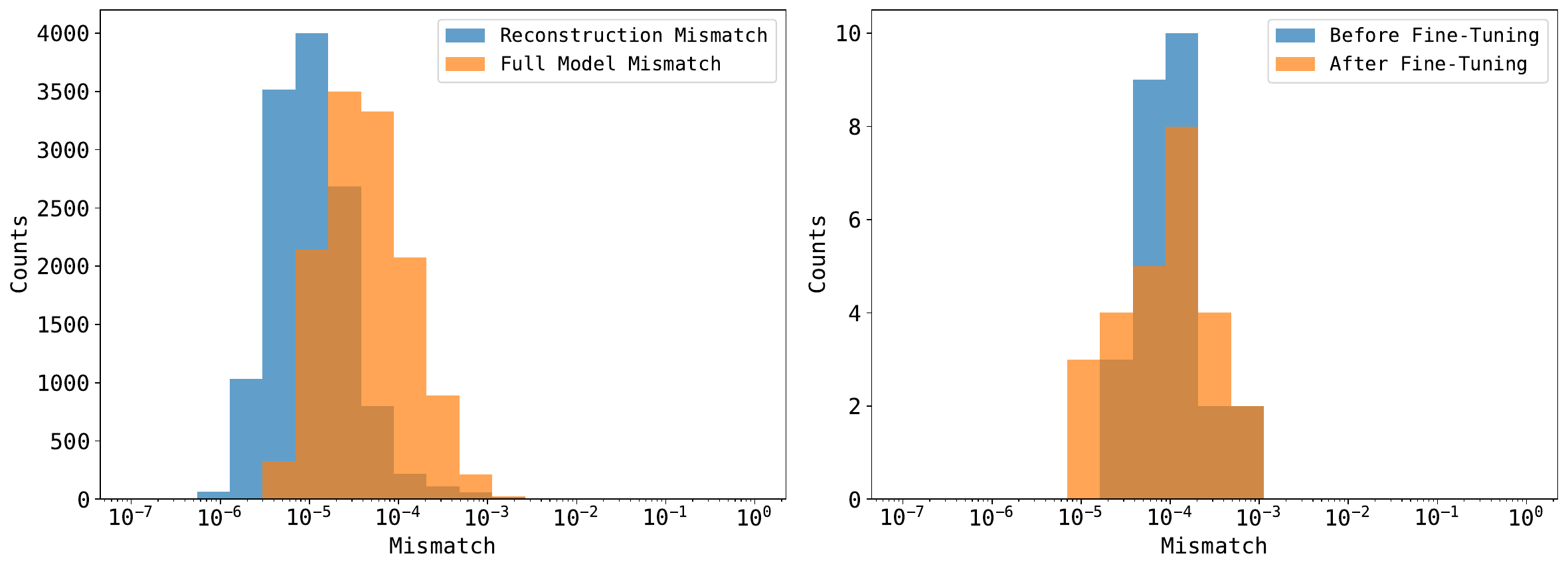}
    \caption{Mismatch distributions for the model-generated waveforms. Left: distribution for the autoencoder reconstruction of the \app test set overlaid with mismatches from waveforms generated by the full model trained on the same dataset. Right: distribution between waveforms generated by the full model and the SXS test set, shown before and after fine-tuning.}
    \label{fig:histograms_accuracy}
\end{figure*}

\begin{table}[htbp]
\centering
\caption{Summary of the key hyperparameters used during the training procedure, where AE, Full, and FT refer to the autoencoder, the full model and the fine tuning stage, respectively.}
\label{tab:hyperparameters_compact}
\begin{tabular}{lc}
\toprule
\textbf{Parameter} & \textbf{Value} \\
\hline
Optimiser & Adam \\
Learning rate & $10^{-4}$ (AE,Full), $10^{-5}$ (FT) \\
Loss function & Mean Absolute Error \\
Batch size & 1024 \\
Training epochs & 3000(AE,Full), 2000 (FT) \\
LR scheduler & ReduceLROnPlateau \\
LR reduction factor & 0.1 \\
LR patience & 250 (AE), 125 (Full, FT) \\
\hline
\hline
\end{tabular}
\end{table}
\subsection{Training}

We start the summary of the training procedure by stating the goal of the final model. We need the model to generate gravitational waveforms for quasi-circular BBH mergers given only their mass ratio and their individual spins, $(q,\chi_1$,$\chi_2)$. Since we observed that training the autoencoder on the amplitude and phase components separately, rather than on the full complex waveform, leads to faster convergence and higher accuracy, the first step in our procedure is to split the dataset into unwrapped amplitudes and phases,
\begin{equation}\label{amplphase}
    h = h_+ -ih_\times = Ae^{i\phi},
\end{equation}
where $h$ is the strain, and $h_+$ , $h_\times$ its components (the ``plus" and ``cross" polarisations), $A$ is the amplitude, and $\phi $ is the phase.
The three stages of the training procedure are as follows:
\begin{enumerate}
    \item 
    We train both autoencoders with the \app dataset; the first one with the amplitudes $A$ and the second one with the phases $\phi$. In this way we obtain the reduced representations we need for the next stages.
    \item Keeping only the decoder part of the resulting autoencoders, we add a fully-connected network to map the physical parameters to the  reduced representations. The ensuing models can predict the amplitude and phase of a waveform given the initial parameters of the system.
    \item  The last stage entails the fine tuning of the full models, which is performed by retraining them with the Numerical Relativity dataset using a smaller initial Learning Rate (LR).
\end{enumerate}
The first two stages take 3000 epochs to complete while the fine-tuning stage 2000 epochs since it stopped improving earlier.
We use the Adam optimiser with an initial $10^{-4}$ LR for the first two stages.
For the fine-tuning stage, as it is commonly done, we choose a smaller $10^{-5}$ LR. We also employ a LR scheduler (\texttt{ReduceLROnPlateau}) which checks the validation loss for each epoch. If it does not improve for a specified number of epochs (LR patience), it reduces the LR by a specified factor (LR reduction factor). 

The evolution of the loss functions during each training stage and for each model is displayed in Fig.~\ref{fig:loss_hist}. We can see that the training concluded with the models having converged in all stages. We do not observe any evidence of overfitting, which would typically be indicated by a validation loss significantly higher than the corresponding training loss, except during the fine-tuning stage of the amplitude model. Even in this case, the difference between the two losses appears to be negligible, as confirmed by the model's performance on the test dataset.

Also, the use of a LR scheduler is justified by the improvement of the loss functions at each step. It is worth noting that, during the fine-tuning stage, we experimented with k-fold cross-validation and weight averaging across the models trained on each fold, following the approach used in our previous work~\cite{GramaxoFreitas:2024bpk}. However, this strategy did not lead to any improvement in performance. Each epoch takes 1 second to complete when the model is trained with the approximant dataset and with a batch size of 1024 on an NVIDIA RTX 4090 GPU. The fine tuning epochs took much less time, in the order of milliseconds, which is expected from the number of available waveforms. A summary of the model hyperparameters is reported on Table~\ref{tab:hyperparameters_compact}.

 \begin{figure}
    \centering
    \includegraphics[width=1\linewidth]{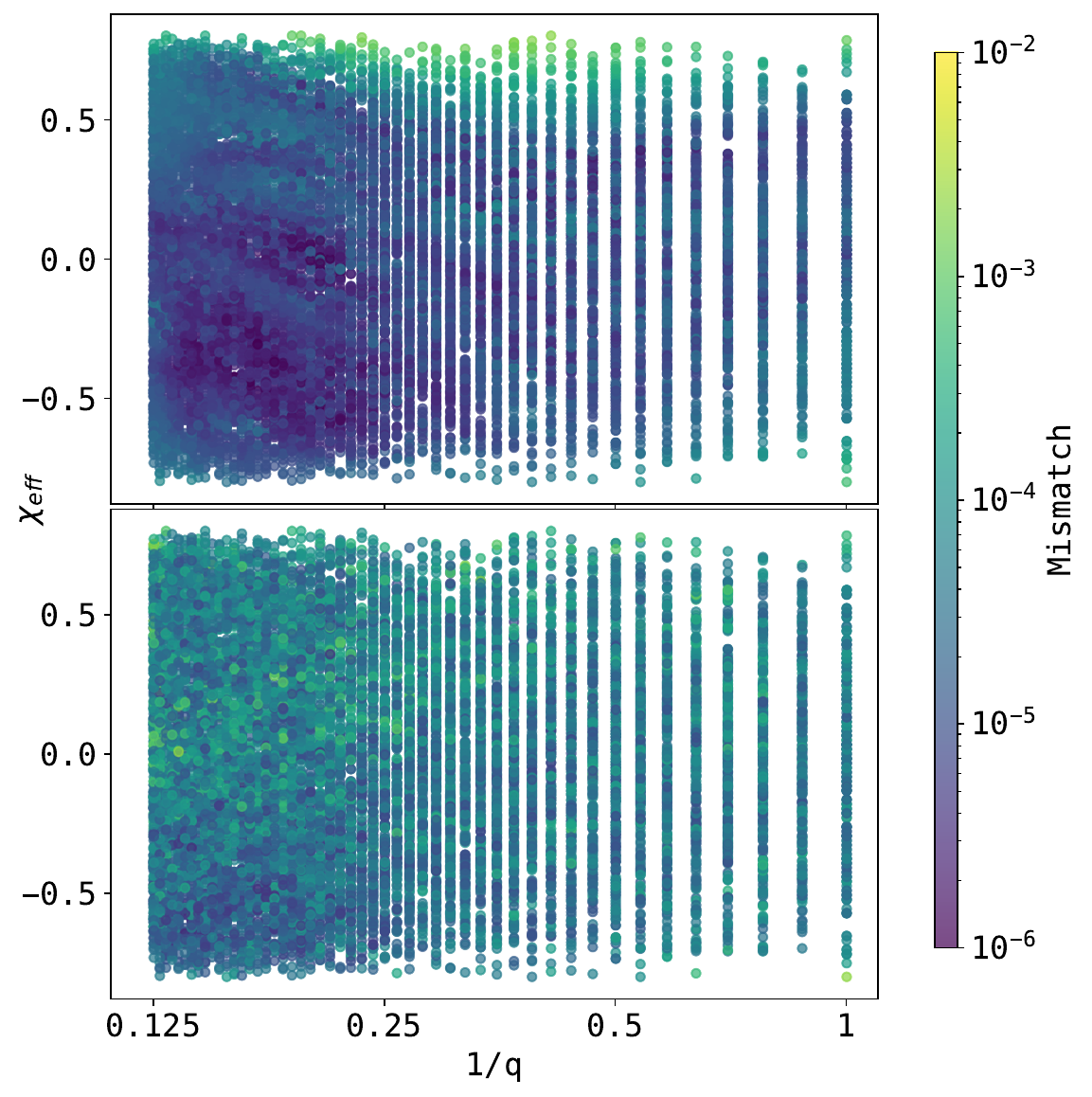}
    \caption{Scatter plot of the distribution of the physical parameters of the \app test dataset, with the spins combined into the effective spin $\chi_{\rm eff}=\frac{\chi_1+q\chi_2}{1+q}$ and coloured with the mismatch between the predicted waveform and the real one, corresponding to each parameter combination. The top panel displays the autoencoder’s reconstruction mismatches at the conclusion of the first training stage, whereas the bottom panel shows the mismatches for the full model at the end of the second stage for the waveforms of the \app test dataset.}
    \label{fig:chieffcomb}
\end{figure}

\section{\label{sec:res} Results}

We turn now to discuss our results when testing the accuracy of our autoencoder-based model inside its training parameter space, along with the time they take to make an inference, and how this time scales with the number of points. We use \textsc{bilby} for the parameter estimation tests~\cite{ashton_bilby_2019}. An inference is carried out for an illustrative BBH gravitational-wave signal detected by the LVK collaboration. To assess the accuracy of our model we employ the mismatch $\mathcal{M}$, a metric commonly used in the literature for the assessment of similarity between generated and real waveforms. The mismatch is related to the overlap $\mathcal{O}$ by
\begin{equation}
  \mathcal{M} (h,~\hat{h}) = 1 -~\mathcal{O}(h,\hat{h}) =1-~\frac{\langle h |~\hat{h}~\rangle}{\sqrt{\langle h | h~\rangle~\langle~\hat{h} |~\hat{h}~\rangle}}\,,
 ~\label{eq:overlap_gravitational_waveforms}
\end{equation}
 where 
\begin{equation}
    \langle h |~\hat{h}~\rangle = 4\,\text{Re}\!\!\int_{0}^{\infty}\!\!\!\! \tilde{h}(f)\tilde{\hat{h}}^*(f) df \approx 4\,\text{Re}\sum_{k}\tilde{h}_k\tilde{\hat{h}}_k^*\Delta f\,
\end{equation}

\begin{figure*}
     \centering
         \includegraphics[width=\textwidth]{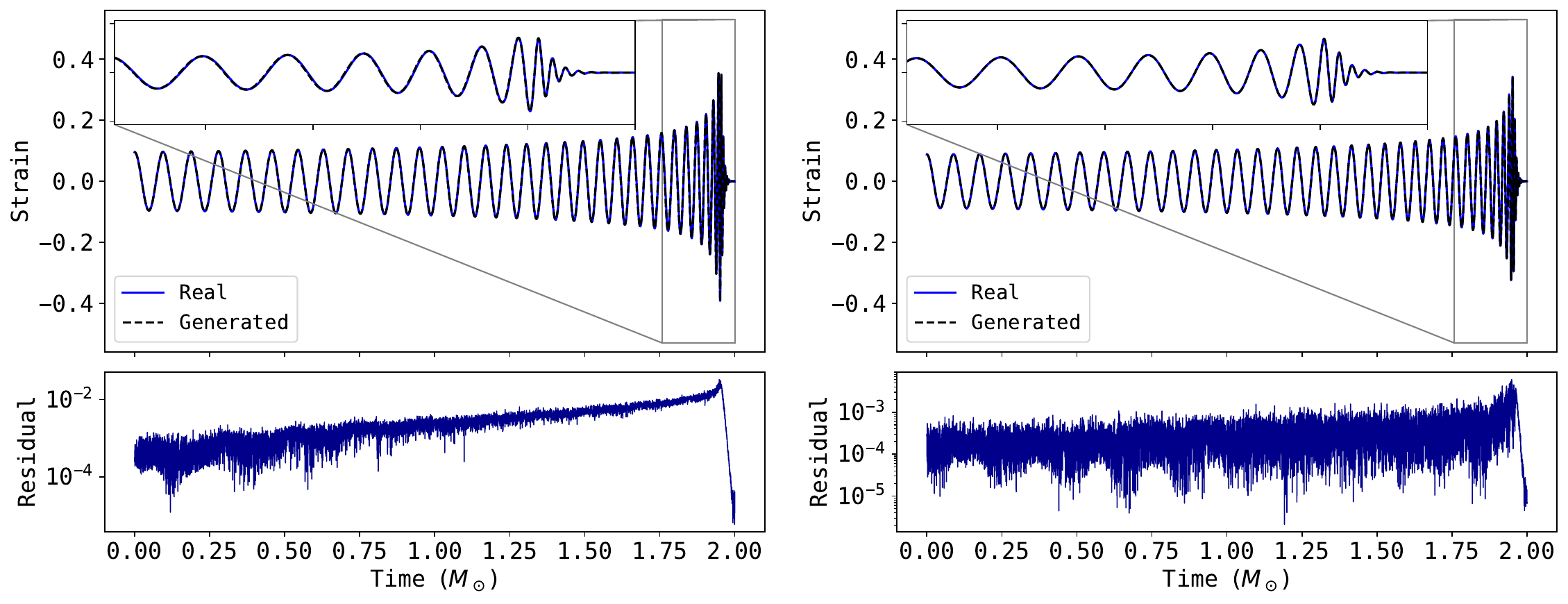}
         \caption{Visualisation of waveforms generated by our model for the SXS test set, overlaid with the corresponding numerical-relativity waveforms to illustrate the model’s performance. Even under close inspection (see insets), the generated waveforms are visually indistinguishable from the true waveforms. The left panel shows the waveform with the highest mismatch in the test set, $\mathcal{M}=7.64\times 10^{-4}$ with input parameters: $1/q = 1.000, \chi_1 =  - 0.2002
         , \chi_2 = - 0.4000$, while the right panel shows the waveform with the lowest mismatch, $\mathcal{M}=1.07\times 10^{-5}$ with input parameters: $1/q = 0.5000, \chi_1 = -0.0001 
         , \chi_2 = 0.0000$. The lower subpanels display the residuals between the generated and true waveforms.
         }
         \label{fig:worst_mismatch}
\end{figure*}

\subsection{Accuracy}

For every waveform in the approximant test set we generate a waveform using our model and its initial parameters. Then, using Eq.~(\ref{eq:overlap_gravitational_waveforms}), we calculate the similarity between them. We repeat the same process for our test split of the SXS dataset before and after we fine-tune our model. The results of these tests are summarised in Table~\ref{tab:Results_accuracy_table},for completeness the PCA-model results are also included \cite{GramaxoFreitas:2024bpk}, and are displayed in the histograms shown in Fig.~\ref{fig:histograms_accuracy} and in the scatter plot of Fig.~\ref{fig:chieffcomb}. 

The mean mismatch obtained for the approximant dataset is of the order of $10^{-5}$ while for the SXS dataset the accuracy degrades by about a factor of 2. The maximum mismatch values attained are associated with the non-negligible tails seen in the histograms on the left of Fig.~\ref{fig:histograms_accuracy}.
This is also illustrated in Fig.~\ref{fig:chieffcomb} which shows the correlation of the physical parameters of the binary with the mismatch between the true and predicted waveforms from the \app test dataset, highlighting regions of significant mismatch in the parameter space. There seems to be an area with higher mismatches, correlated with the tails in the left histograms, for higher effective spin. This high mismatch area does not persist after the training of the full model and does not correspond to the one present in our PCA-based model (see Fig.~7 in~\cite{GramaxoFreitas:2024bpk}).
This can be attributed to difference of sampling distribution and the higher expressivity and nonlinearity of our autoencoder model compared to PCA.
The fine-tuning stage appears to change marginally the accuracy of the model, since it reduces the impact of the high mismatch tail by 3.2\% while, at the same time, slightly increases the mean mismatch by 1.3\%. We find the impact on the higher mismatches to be more important than the minor decrease of the mean and, thus, we keep the fine tuned model as the one to be used further. 

Overall, our model is accurate enough to be used as a surrogate waveform generation model. This is illustrated in Fig.~\ref{fig:worst_mismatch} where we display the waveform with the worst mismatch in the SXS dataset after fine-tuning,$\mathcal{M}=7.64\times 10^{-4}$, along with the one with the best mismatch. No significant differences are observed even when zooming in in the merger part of the signal. The lower panels of the same figure show the amplitude residuals between the actual and generated waveforms. In both cases, the residuals increase toward merger, where the amplitude varies more rapidly. This effect is more pronounced in the left lower panel which corresponds to the waveform with the highest mismatch. 
Finally, we observe that \dan performs $7\times$ better on average, while our fine-tuned model (\aes) outperforms our previous model when it comes to the highest mismatch by attaining a $3.5\times$ lower value.
\begin{table}[htbp]
\centering
\caption{Summary of the accuracy of our model during the different stages of training and with the two datasets used. Both the mean and the maximum values of the mismatch are reported. We have also added the results of our PCA-model, for completeness.}
\label{tab:Results_accuracy_table}
\centering
\begin{tabular}{cccccc}
\hline
\hline

Model & Dataset & Mean & Max  \\
\hline
Autoencoder & \app & $2.40\times10^{-5}$ & $1.96\times10^{-3}$ \\
Full & \app  & $8.26\times10^{-5}$ & $5.16\times10^{-3}$ \\
Full & SXS  & $1.54\times10^{-4}$ & $7.89\times10^{-4}$ \\
Fine-tuned & SXS  & $1.56\times10^{-4}$ & $7.64\times10^{-4}$ \\
\dan & SXS  & $2.30\times10^{-5}$ & $2.70\times10^{-3}$ \\



\hline
\hline
\end{tabular}

\end{table}

\subsection{Speed}\label{sec:res_speed}

We now discuss the time it takes the model to generate waveforms for different number of samples. In order to measure this time we take the following steps:
\begin{enumerate}
    \item Compile the model by doing a computation with a small number of samples.
    \item Perform 3 warm-up computations with an actual sample size, to get rid of ``outliers" from the first computation.
    \item Perform 10 computations with different number of samples. 
\end{enumerate}
In this way, we ensure that neither model initialisation nor random time delays affect the speed measurement, yielding an unbiased assessment of the model’s performance in waveform generation and its scalability with the number of samples. We also make use of batches, since as the number of samples increases the weights and the outputs of the neurons do not fit any more in GPU memory and the model fails to run. With the use of a batch size of min(10000, Number of samples) we make sure that the models run correctly with our available GPU memory (NVIDIA RTX 4090). Having higher usable GPU memory would allow running larger batches and thus speeding up the total waveform generation time, as the forward pass itself is a very rapid and parallelisable task on GPUs.


Our results are summarised in Table~\ref{tab:timing_table}.
We benchmarked our model using its serialised just-in-time (JIT) version, comparing it with a rerun of the JIT version of \dan on the same hardware and \app run via the \gwsur package. It is important to note that our models' raw timings correspond to a single sequential time-domain forward pass of the two networks, which output the $(2,2) $ amplitude and phase, respectively, on a fixed time grid in natural units (sampling rate $2 M_\odot^{-1} $, window $4096 M_\odot$) as well as the construction of the waveform using Eq.\ref{amplphase}. For a fair comparison, we time \app generating only the (2,2) mode on a 8192 time grid like our autoencoders output (mode\_list=[(2,2)], dt = 0.5 in geometric units).  We also note that \app is run sequentially, as parallel execution on 32 CPU cores resulted in slower performance. The other two models seem to leverage better the higher number of cores of our AMD Ryzen Threadripper PRO 5975WX CPU, so these times are reported in Table~\ref{tab:timing_table}.
\\
We find that even the CPU implementation of our model remains significantly faster than \texttt{NRHybSur3dq8}, making it suitable for low-latency parameter estimation. GPU acceleration provides additional gains as the number of samples increases managing to generate 1 million waveforms in under a second. Finally, it is worth pointing out that our PCA-based surrogate~\cite{GramaxoFreitas:2024bpk} substantially outperforms \texttt{AESur3dq8}, generating 1 million waveforms 5 times faster (in 172 ms), primarily due to the simplicity of the inverse PCA transform compared to the more complex decoder network. Another factor adding to that, is the higher output size of the autoencoder, 8192 compared to the 2048 of \dan, as it substantially increases the number of parameters in the model.

\begin{table}[h]
    \centering
    \caption{
    Mean waveform generation time (ms) for \aes and \dan  on CPU and GPU across different sample sizes, compared with \app run via the \textsc{gwsurrogate} package. The values with an asterisk are extrapolated using the sample/s values from the previous computation as it does not vary with time.   
    }
    \label{tab:timing_table}


\begin{tabular}{cccccc}
\hline\hline
 & \multicolumn{2}{c}{\aes
 }  & \multicolumn{2}{c}{\dan}& \multicolumn{1}{c}{\app} \\
\hline
Samples & CPU  & GPU  &CPU  & GPU &CPU   \\
\hline
$10^{2}$ & 6.38    & 0.46 & 1.0    & 0.18 & 881 \\
$10^{3}$ & 38.0     & 1.01 & 3.0    & 0.21 & 8,830 \\
$10^{4}$ & 349   & 9.05  & 125    & 1.8  & 88,200 \\
$10^{5}$ & 3,682  & 86.6   & 1,222  & 17   & $882,000^*$ \\
$10^{6}$ & 37387 & 868  & 12,217 & 172  & $8.8 \times 10^{6*}$ \\
\hline\hline
\end{tabular}




\end{table}




\subsection{Parameter estimation}

Having established the accuracy of the surrogate waveforms generated with our autoencoder-based model we turn now to test its capabilities for gravitational-wave parameter estimation. To do so we use \gwsur, which takes care of the extrinsic parameters and unit conversions and we employ Bayesian inference with nested sampling \cite{original_nested_sampling}, the most common algorithm for parameter estimation in gravitational-wave data analysis. Nested sampling relies in repeatedly evaluating a likelihood function that requires waveforms. Usually, these waveforms are obtained from surrogate models or approximants, and the speed of the waveform generation stage affects greatly the overall speed of the parameter estimation algorithm.
As discussed previously, our model shows a significant speedup compared with \app so we expect a speedup in parameter estimation via nested sampling when using \aes. However, it is important to note that \gwsur and \textsc{bilby} do not currently support GPU acceleration, to the best of our knowledge, so the model inference is done on CPU and, thus, we expect the final speedup to be modest.  

We conduct this test by performing two parameter estimation runs with the \textsc{bilby} (v2.7.1) Bayesian inference package \cite{ashton_bilby_2019} and the \textsc{nessai} (v0.15) sampler \cite{nessai}, one using the \app surrogate model and the other using our method. For computational efficiency, we stopped the sampler once the increase in the log-evidence fell below a threshold of one. Since \textsc{bilby} is built on CPUs we cannot leverage the speedup provided by parallel computing for assessing the benefits of using our model. Hence, all tests presented in this section were performed on CPUs, specifically on an Intel Xeon Platinum 8160 processor running at 2.10 GHz with 24 cores. For its usage in \textsc{bilby}, our fine-tuned neural network was compiled using JIT compilation and wrapped using \textsc{gwsurrogate} (v1.1.8) for projecting the waveform in the sky and rescaling it in SI units. Then, we ran the sampler on the real strain data containing the GW200311\_115853 event from both of LIGO's interferometers data along with Virgo data. This event was selected to allow a direct comparison with the results presented in \cite{GramaxoFreitas:2024bpk} using our PCA-based approach. 

\begin{figure*}
    \centering
    \includegraphics[width=0.95\linewidth]{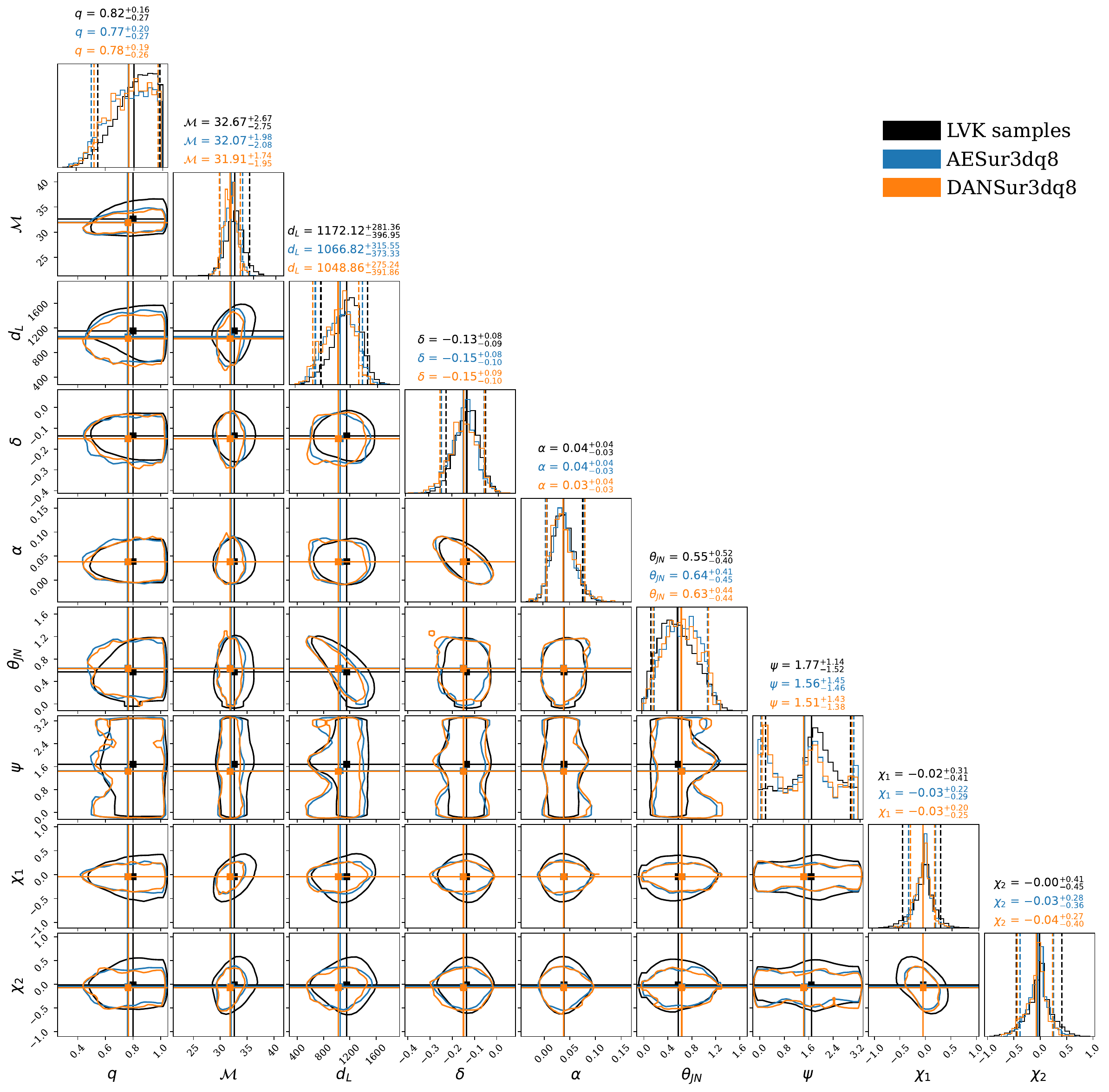}
    \caption{Results comparison between a parameter estimation run using our model and the results published by the LVK collaboration for the GW200311\_115853 event. The published results are outlined in black, our results are represented in orange and the results obtained using \dan, our PCA-based model, are shown in blue. The magnitudes estimated are, from left to right, the binary mass ratio, the chirp mass, the luminosity distance, the sky localisation angles (declination and right ascension), orbital inclination, polarisation angle and the spin $z$-component magnitudes for each object. All estimated medians of the distributions obtained with our model lie within the LVK uncertainty bounds and the obtained distributions correlate fairly well visually. Our results present some multimodality in the polarisation angle, probably due to the fact that our model only produces the $l=m=2$ multipole. We would expect this multimodal distribution to collapse into a unimodal one when higher order multipoles are included in the model.}
    \label{fig:pe_bilby}
\end{figure*}

\begin{table}[t]
    \centering
    \caption{Computational cost of parameter estimation using
    \texttt{AESur3dq8} and \app. Values correspond to the
    mean and standard deviation over ten independent runs. The speed-up is
    defined as the ratio between the \app and
    \texttt{AESur3dq8} mean timings, such that values larger than unity
    indicate an acceleration obtained with \texttt{AESur3dq8}.
    Here, $t_{\rm samp}$ denotes the total sampling time,
    $t_{\mathcal{L}}$ the cumulative time spent evaluating the likelihood,
    and $N_{\mathcal{L}}$ the total number of likelihood evaluations.}
    \label{tab:pe_timings}
    \small
    \setlength{\tabcolsep}{4pt}
    \begin{tabular}{lccc}
        \hline\hline
        Quantity
        & \texttt{AESur3dq8}
        & \app
        & Speed-up \\
        \hline
        $t_{\rm samp}$ (h)
        & $0.557 \pm 0.107$
        & $1.116 \pm 0.204$
        & $2.00\times$ \\

        $t_{\mathcal{L}}$ (h)
        & $0.285 \pm 0.069$
        & $1.053 \pm 0.200$
        & $3.69\times$ \\

        $t_{\mathcal{L}}/N_{\mathcal{L}}$ (ms)
        & $1.378 \pm 0.307$
        & $19.08 \pm 2.17$
        & $13.8\times$ \\

        $N_{\mathcal{L}}$ ($10^{5}$)
        & $7.44 \pm 0.65$
        & $1.98 \pm 0.20$
        & --- \\
        \hline\hline
    \end{tabular}
\end{table}

An overlapping corner plot showing the obtained results on top of each other is provided in Fig.~\ref{fig:pe_bilby}. We can see that the medians obtained by our run lie well within the confidence intervals published by the LVK collaboration, while the distribution shapes visually correlate fairly well with the exception of the polarisation angle, which presents some multimodality which is not present in the LVK samples. This is probably due to the fact that our model only produces the dominating multipole, leaving out crucial information that would break the multimodality in the polarisation angle. We also observe that the inferred parameters are distributed very similarly when using either of the two approaches, based on PCA or autoencoders.

As discussed in section \ref{sec:res_speed}, our model provides a significant speedup in terms of waveform generation, which should result in accelerating likelihood-based inference algorithms. To test this we ran ten independent inference runs with the same characteristics to compare the algorithm speed when using \app compared with \aes and compiled the logged timing results, which are shown in table \ref{tab:pe_timings}. As we can see, for this particular example, the \aes runs need around $\sim3.76$ times more likelihood evaluations to converge. We attribute this to the difference in waveform length but this is something we will investigate further in future work.
Despite this, the total inference time is $\sim3.69$ times lower due to the fact that the time spent per likelihood evaluation is much faster $\sim13.8$. 


\section{\label{sec:con} Conclusions}

Machine-learning–based surrogate waveform models have been shown to significantly accelerate waveform generation while achieving levels of accuracy comparable to those of traditional waveform approximants (see e.g.~\cite{GramaxoFreitas:2024bpk}). In this paper we have investigated the viability of artificial neural networks with an autoencoder-based training strategy to generate an efficient reduced representation of gravitational-wave signals from quasi-circular BBH mergers. This work extends our previous study~\cite{GramaxoFreitas:2024bpk}, in which we performed a similar assessment using a neural-network–based surrogate model built upon a reduced PCA basis of approximant waveforms. 

Our autoencoder-based surrogate waveform model \aes generates the (2,2) mode of gravitational wave signals from \qcbbh 
over its final 8192 samples ($4096M_\odot$ at $\Delta t = 0.5M_\odot$) for $q\in [1,8]$ and $\chi_1,\chi_2 \in [-0.8,0.8]$ and is capable of rapidly generating large template banks, producing millions of waveforms in under a second on GPUs using modest computational resources. Even on standard CPUs, \aes can perform the same task in under 40 seconds. The model training relies on a three-stage procedure. As a first step, we train an autoencoder on the \app approximant dataset to obtain a reduced, low-dimensional representation of the waveforms. Next, we repurpose the resulting decoder by introducing a mapping network before its input and train the combined system as a single network, again using the \app dataset. This procedure transforms the autoencoder into a hybrid artificial neural network that can generate gravitational-wave signals from quasi-circular BBH systems. Finally, the model is fine-tuned using the SXS Numerical Relativity dataset, which 
enhances its accuracy, only marginally. The fine-tuning stage reduces the worst-case mismatch by about 3\% while increasing the mean mismatch by 1.3\%.  Lastly, comparing with our \dan model, we find that \dan is about $5\times$ faster on GPU and $3\times$ faster on CPU (for $10^6$ samples), and roughly $7\times$ more accurate in terms of mean mismatch. The autoencoder model, however, attains a worst-case mismatch about $3.5\times$ lower than \dan. The PCA-based model appears better suited to this dataset, while the autoencoder-based one is better suited for applications where a more robust worst-case behaviour is needed.
In the last part of this paper, we have shown that the parameter inference performed using templates generated by \aes leads to fully consistent results with those reported by the LVK Collaboration and with those using our PCA-based model~\cite{GramaxoFreitas:2024bpk} for the gravitational-wave event GW200311\_115853, used as an illustrative example. 


\vspace{0.2cm}
\begin{acknowledgments}
AT acknowledges support from the Universitat de València through the Atracció de Talent fellowship.
NVE is supported by the Valencian Government Grant No. CIAICO/2024/111 and the Spanish Ministry of Economic Affairs and Digital Transformation through the QUANTUM ENIA project call – Quantum Spain project, and the European Union through the Recovery, Transformation and Resilience Plan – NextGenerationEU (Digital Spain 2026) Agenda.
OGF is supported by the Portuguese Foundation for Science and Technology (FCT) through doctoral scholarship UI/BD/154358/2022. 
TF is supported by FCT through doctoral scholarship 2023.03753.BD. 
SN is supported by FCT through doctoral scholarship 2025.02005.BD.
OGF, TF, SN and AO acknowledge financial support by CF-UM-UP through the Strategic Fundings UIDB/04650/2020, UIDP/04650/2020, UID/PRR/04650/2025 and UID/04650/2025 of FCT.
JAF and ATF are supported by the Spanish Agencia Estatal de Investigaci\'on (grant PID2024-159689NB-C21) funded by MICIU/AEI/10.13039/501100011033 and by FEDER/EU, by the Generalitat Valenciana (Prometeo grant CIPROM/2022/49), and by the European Horizon Europe staff exchange  (SE)  programme HORIZON-MSCA-2021-SE-01 (grant NewFunFiCO-101086251). 
JDMG is partially supported by the agreement funded by the European Union, between the Valencian Ministry of Innovation, Universities, Science and Digital Society, and the network of research centers in Artificial Intelligence (Valencian Foundation valgrAI), as well as the Valencian Government grant with reference number CIAICO/2024/111; the Spanish Ministry of Economic Affairs and Digital Transformation through the QUANTUM ENIA project call – Quantum Spain project, and the European Union through the Recovery, Transformation and Resilience Plan – NextGenerationEU within the framework of the Digital Spain 2025 Agenda. 
The authors gratefully acknowledge the computer resources at Artemisa and the technical support provided by the Instituto de Fisica Corpuscular, IFIC (CSIC-UV). Artemisa is co-funded by the European Union through the 2014-2020 ERDF Operative Programme of Comunitat Valenciana, project IDIFEDER/2018/048. 
This material is based upon work supported by NSF's LIGO Laboratory which is a major facility fully funded by the National Science Foundation, as well as the Science and Technology Facilities Council (STFC) of the United Kingdom, the Max-Planck-Society (MPS), and the State of Niedersachsen/Germany for support of the construction of Advanced LIGO and construction and operation of the GEO600 detector. Additional support for Advanced LIGO was provided by the Australian Research Council. Virgo is funded, through the European Gravitational Observatory (EGO), by the French Centre National de Recherche Scientifique (CNRS), the Italian Istituto Nazionale di Fisica Nucleare (INFN) and the Dutch Nikhef, with contributions by institutions from Belgium, Germany, Greece, Hungary, Ireland, Japan, Monaco, Poland, Portugal, Spain. KAGRA is supported by Ministry of Education, Culture, Sports, Science and Technology (MEXT), Japan Society for the Promotion of Science (JSPS) in Japan; National Research Foundation (NRF) and Ministry of Science and ICT (MSIT) in Korea; Academia Sinica (AS) and National Science and Technology Council (NSTC) in Taiwan. 
\end{acknowledgments}


\bibliography{Bibliography/Parameter_estimation,Bibliography/Ligo,Bibliography/Surrogates,Bibliography/ML,Bibliography/Websites,Bibliography/Others,Bibliography/dictionaries}

\end{document}